**Total variation superiorization schemes in proton computed tomography image reconstruction**

S. N. Penfold[a]

5   Centre for Medical Radiation Physics, University of Wollongong, Wollongong, NSW, Australia

R. W. Schulte

Department of Radiation Medicine, Loma Linda University Medical Center, Loma Linda, CA, USA

Y. Censor

Department of Mathematics, University of Haifa, Mt Carmel, Haifa, Israel

10  A. B. Rosenfeld

Centre for Medical Radiation Physics, University of Wollongong, Wollongong, NSW, Australia

[a]Corresponding author. Phone +61 4221 8194; Email: snp75@uow.edu.au

## ABSTRACT

**Purpose:** Iterative projection reconstruction algorithms are currently the preferred reconstruction method in proton computed tomography (pCT). However, due to inconsistencies in the measured data arising from proton energy straggling and multiple Coulomb scattering, noise in the reconstructed image increases with successive iterations. In the current work, we investigated the use of total variation superiorization (TVS) schemes that can be applied as an algorithmic add-on to perturbation-resilient iterative projection algorithms for pCT image reconstruction.

**Methods:** The block-iterative diagonally relaxed orthogonal projections (DROP) algorithm was used for reconstructing Geant4 Monte Carlo simulated pCT data sets. Two TVS schemes added on to DROP were investigated; the first carried out the superiorization steps once per cycle and the second once per block. Simplifications of these schemes, involving the elimination of the computationally expensive feasibility proximity checking step of the TVS framework, were also investigated. The modulation transfer function and contrast discrimination function were used to quantify spatial and density resolution, respectively.

**Results:** With both TVS schemes, superior spatial and density resolution was achieved compared to the standard DROP algorithm. Eliminating the feasibility proximity check improved the image quality, in particular image noise, in the once-per-block superiorization, while also halving image reconstruction time. Overall, the greatest image quality was observed when carrying out the superiorization once-per-block and eliminating the feasibility proximity check.

**Conclusions:** The low contrast imaging made possible with TVS holds a promise for its incorporation into our future pCT studies.

## I. INTRODUCTION

Proton computed tomography (pCT) has been suggested as a means for reducing the range uncertainty in proton radiation therapy[1]. As demonstrated in previous work[2,3], iterative algebraic techniques incorporating a most likely path formalism[4] are required to improve spatial resolution and quantitative accuracy of pCT image reconstruction. Analysis of image quality measures when reconstructing simulated pCT data with iterative techniques has demonstrated that both spatial resolution and mean reconstructed relative stopping power (RSP) values improve with an increasing number of iterations. However, amplification of image noise through the iterative procedure limits the monotonic increase in image quality with iteration number.

The iterative algorithmic schemes presented in previous pCT image reconstruction work[2,3,5] belong to the general class of feasibility seeking methods[6,7]. That is, the algorithm searches for a solution, and there may be many solutions, to the convex feasibility problem (CFP) of finding a point in the intersection of a finite family of convex sets. This is different from optimization, which seeks a solution to the problem statement by minimizing a given cost (merit) function over the constraints sets of the CFP.

While optimization has certain advantages in many imaging applications, it could have drawbacks in applications such as pCT. The optimal solution, in a mathematical sense as dictated by the cost function, may not always be the solution that best reproduces the true object data of interest, because of inconsistencies in the acquired data or due to the choice of the cost function. This choice is affected by justifying arguments that are sometimes inadequate or by the ability or inability to computationally handle the resulting optimization problem. Therefore, in this work we investigated the potential value of the superiorization method, which is also attractive both in terms of required memory and computational time.

In the current work, we investigate the superiorization of total variation (TV). While TV-minimization has been used extensively as a denoising tool in the field of image processing[8,9,10], TV-superiorization (TVS) is a new methodology in image reconstruction. The concept was
65 introduced by Butnariu *et al.*[11] (although the term TV-reducing is used there and not superiorization as used here). There, the authors proved the perturbation resilience, i.e. stability, of string-averaging projection (SAP) methods under summable perturbations, and proposed how to use this resilience to steer a feasibility seeking iterative process toward iterates with reduced TV values. The usefulness of this method when applied to X-ray CT reconstruction from a limited number of projections was
70 demonstrated by Herman and Davidi[12]. Perturbation resilience of block-iterative projection (BIP) methods was later presented by Censor, Davidi, and Herman[13].

The premise is that the problem at hand (image reconstruction in pCT in our case) is modeled by a CFP, but that we desire (i) to use an efficient feasibility seeking projection method, and (ii) to find a feasible solution that will have a reduced value of a given merit function (TV in our case).
75 Superiorization refers to such a process of finding a superior solution with respect to some merit function, which is also a feasible solution of the CFP sets. A superior solution is a feasible solution of the CFP for which the value of the merit function, with respect to which one superiorizes, is smaller (but not necessarily minimal) than the value of this function at the feasible point that would have been reached if the superiorization process would not have been applied. The ability to perturb
80 the original projection algorithm, without losing convergence to a feasible point, allows us to steer the algorithm toward a feasible point that is superior, according to the merit function, to the one we would arrive at without the perturbations.

Superiorization is fundamentally different from constrained minimization. The novelty lies in the attempt to strike another balance between feasibility and minimality. The term superiorization
85 reflects the main idea of the new approach, which is not the finding of any feasible point (solving the feasibility problem) and not the quest for a constrained minimum point. Instead, the target of superiorization is to seek a feasible point that is also "better", i.e., superior in a defined mathematical sense, over other reachable feasible points with respect to the objective function. Such a point does not need to be a minimum point of the objective function in the feasible set but its superiority means that it is, in some rigorously well-defined sense, "more than" just a feasible point.

We first give a general formulation of a prototypical algorithmic framework which is based on, but not identical with, the general framework for superiorization given in Ref. 14. Then, we use Monte Carlo simulated data and a quantitative analysis of image quality, to investigate the application of two TVS schemes to pCT image reconstruction. Both TVS schemes are constructed by modifying the block-iterative diagonally relaxed orthogonal projections (DROP) algorithm[15] as the core reconstruction algorithm. The first scheme employs the superiorization steps *once per cycle*, where a cycle is a complete processing of all data blocks. The second TVS scheme employs the superiorization steps *once per block*. Simplifications of these schemes were also investigated in which the computationally expensive feasibility proximity checking step of the TVS framework was eliminated.

Our main conclusion is that superiorization is a useful reconstruction scheme for pCT that can be applied as an algorithmic add-on to perturbation-resilient iterative projection algorithms seeking feasibility. In addition, there are significant advantages of the TVS schemes in detecting small contrast differences in pCT images.

## II. METHODS

### A. The superiorization methodology

The superiorization principle has its roots in Ref. 11 and was recently formalized in Ref. 14.

It relies on the notion of bounded perturbation resilience of algorithms. An *algorithm* **P** is said to be *resilient to bounded perturbations* if the following is the case: if the sequence $\left(\left(\mathbf{P}\right)^k x\right)_{k=0}^{\infty}$ (obtained by sequential repeated applications of **P**, starting from $x$) converges to a solution of *problem Q* for all $x \in R^n$, then any sequence $\left(x^k\right)_{k=0}^{\infty}$ of points in $R^n$ also converges to a solution of $Q$ provided that, for all $k \geq 0$,

$$x^{k+1} = \mathbf{P}_Q\left(x^k + \beta_k v^k\right), \qquad (1)$$

where $\beta_k v^k$ are bounded perturbations, meaning that $\beta_k$ are real nonnegative numbers such that $\sum_{k=0}^{\infty} \beta_k < \infty$ and the sequence of vectors $\left(v^k\right)_{k=0}^{\infty}$ is bounded.

The aim of the superiorization methodology is to handle models represented by a constrained minimization problem differently. Instead of trying to solve a constrained minimization problem, it proposes to perturb some powerful feasibility seeking algorithms so that, without losing their convergence toward feasibility, they will yield a point (or points) with reduced objective function value(s). To develop and apply a superiorization approach to some problem we need to secure perturbation resilience of the "powerful" algorithm at hand and then interlace into its steps admissible perturbations that will in some way reflect and help reduce the objective function values.

Specific instances of algorithms resilient to bounded perturbations for solving the convex feasibility problem from the classes of string-averaging projections (SAP) and block-iterative projections (BIP) methods were presented in Ref. 14. Relying on the mathematically validated notion of bounded perturbations resilience, the superiorization theory is currently a heuristic, but practical demonstrations of its usefulness, see Refs. 11,12,13 and 14, are reinforced by our present investigation.

The superiorized algorithms, which are described below, are close to the ones in Refs. 11, 12
and 13, but are different from the specific superiorized algorithm investigated in Ref. 14. The different superiorized algorithms are characterized by the frequency of the perturbation within a cycle over the collected data, and by where exactly in the algorithmic flow the merit function value check and the feasibility proximity check are done. Also, we have investigated the effect of eliminating the computationally expensive feasibility proximity check altogether. Full details about our specific implementations appear in the following section. These implementations are concentrated solely around the use of superiorized algorithms for total variation reduction in pCT. But a general description of the superiorized algorithm is general enough to apply to other inverse problems in which a reduction (which is not necessarily a minimization) of a given merit function, subject to convex feasibility constraints, is required.

**B. Total variation superiorization applied to proton CT**

In our application of the superiorization scheme to pCT image reconstruction, we have adopted the diagonally relaxed orthogonal projections (DROP) algorithmic scheme[15] for the projection operator **P**, which we have used in previous work[3]. We first give an overview of the iterative reconstruction problem encountered in pCT.

Let $I = \{1, 2, ..., m\}$, and let $\{H_i \mid i \in I\}$ be a finite family of hyperplanes in $R^n$. In pCT reconstruction, the sets $H_i$, on which the vectors $x^k$ are projected during the iterative process, are defined by the $i$-th row of the $m \times n$ linear system $Ax = b$, namely,

$$H_i = \left\{ x \in R^n \mid \langle a^i, x \rangle = b_i \right\}, \quad \text{for} \quad i = 1, 2, ..., m. \quad (2)$$

Here $a^i$ is the $i$-th column vector, of $A^T$ (the transpose of $A$), i.e., its components occupy the $i$-th row of $A$. The right-hand side vector is $b = (b_i)_{i=1}^m$. In pCT, the $a_j^i$ correspond to the length of intersection of the $i$-th proton history with the $j$-th voxel, $x$ is the unknown image vector of relative stopping power (RSP) values, and $b_i$ is the integral RSP along the most likely path (MLP) of the *i*-th proton calculated from its measured energy loss. See Penfold *et al.*[5] for a detailed explanation of the MLP chord length calculation process. The elements $b_i$ are calculated with

$$b_i = \int_{E_{out}}^{E_{in}} \frac{dE}{S_{water}(E)}, \tag{3}$$

where $E_{in}$ and $E_{out}$ are the known entry energy and measured exit energy, respectively, and $S_{water}$ is the proton stopping power of water given by

$$S_{water} = \frac{4\pi r_e^2 m_e c^2 \eta_{water}}{\beta^2(E)} \left[ \ln\left( \frac{2m_e c^2}{I_{water}} \frac{\beta^2(E)}{1-\beta^2(E)} \right) - \beta^2(E) \right]. \tag{4}$$

Here, $r_e$ is the classical electron radius, $m_e$ is the mass of the electron, $\eta_{water}$ and $I_{water}$ are the electron density and mean ionization of potential of water respectively, and $\beta$ is the velocity of the proton relative to the speed of light *c*.

A block-iterative version of DROP with fixed blocks was used by partitioning the indices of $I$ as $I = I_1 \cup I_2 \cup \ldots I_M$ into $M$ blocks. Block-iterative DROP is a variant of the general block-iterative projection method[16], that employs a component-dependent weighting scheme. Block-iterations for the linear case were first studied in Ref. 17. The block-iterative DROP algorithm is given as follows.

**Algorithm 1:** Diagonally Relaxed Orthogonal Projections (DROP)

*Initialization:* $x^0 \in R^n$ is arbitrary.

*Iterative Step:* Given $x^k$, compute the next iterate $x^{k+1}$ with,

$$x^{k+1} = \mathbf{P}_{t(k)}(x^k) = x^k + \lambda_k U_{t(k)} \sum_{i \in I_{t(k)}} \frac{b_i - \langle a^i, x^k \rangle}{\|a^i\|^2} a^i \tag{5}$$

Here, the diagonal matrix $U_{t(k)} = \text{diag}\left(\min\left(1, 1/h_j^t\right)\right)$ with $h_j^t$ being the number of proton histories in the $t$-th block that intersect the $j$-th pixel, and $(\lambda_k)_{k=0}^{\infty}$ is a sequence of user-determined relaxation parameters. In the current work, $\lambda$ was kept at a value of 1.9, based on experience from our previous reconstruction work (unpublished). The blocks are taken up by the algorithm according to the control sequence $(t(k))_{k=0}^{\infty}$ which is in our work a cyclic control, i.e., $t(k)=k$ mod $M+1$. The pCT data set was partitioned into 12 blocks of equal size and composed of an equal number of proton histories from each projection angle. We will refer to this generic DROP, i.e., without superiorization steps added to it, as "standard DROP".

The merit function $\phi$ and feasibility proximity function $Pr$ used in the current work to steer the superiorization reconstruction scheme were motivated by the work of Butnariu et al.[11]. The feasibility proximity function was associated with the residual of measured integral RSP values and those obtained with the current image estimate. The purpose of feasibility proximity checking was to ensure that superiorization with respect to an additional task represented by the merit function $\phi$ did not steer the solution away from an agreement with the measured data. The feasibility proximity of the current image estimate $x^k$ to the measured data was calculated as

$$Pr\left(x^k\right) = \sqrt{\sum_{i=1}^{m}\left(\frac{b_i - \langle a^i, x^k \rangle}{\|a^i\|}\right)^2}. \tag{6}$$

where $m$ is the number of proton histories in the set of interest.

The superiorization paradigm has not yet been investigated for situations where the underlying "feasible set" (of the intersection of the constraints) is empty. But even in such a situation, reducing the proximity function of Eq. (6) is leading to a point which "violates the constraints less", and thus is useful, even if the proximity function does not reach (and cannot reach) the value zero.

The merit function $\phi$, which we aim to reduce during the reconstruction process, was associated with the total variation of the reconstructed image estimate, such that

195
$$\phi(p^k) = \sum_{g=1}^{J-1} \sum_{l=1}^{J-1} \sqrt{\left(p_{g+1,l}^k - p_{g,l}^k\right)^2 + \left(p_{g,l+1}^k - p_{g,l}^k\right)^2}, \quad (7)$$

where $p^k$ is the 2D $J \times J$ representation of the $n$-dimensional image vector $x^k$.

Finally, the perturbation vectors $v^k$, steering the iterative sequence of image estimates toward reduced total variation of the image estimate, were calculated with the method proposed in Ref. 11. Specifically, the perturbation vector was calculated as the negative of the normalized subgradient of
200 the total variation at $x^k$, i.e.,

$$v^k = \begin{cases} -\dfrac{s^k}{\|s^k\|}, & \text{if } s^k \neq 0, \\ 0, & \text{otherwise.} \end{cases} \quad (8)$$

The subgradient of total variation, $s$, was calculated with the method outlined in Ref. 8.

Two variants of the DROP based superiorization scheme, TVS1-DROP (Algorithm 2) and TVS2-DROP (Algorithm 3), were employed in the current work, essentially differing in the number
205 of times the projection operator **P** was applied before continuing to the feasibility proximity check. In both variants, the initial image estimate of the iterative procedure was acquired by performing a filtered backprojection (FBP) reconstruction from the data. The FBP was carried out by rebinning individual proton histories, to conform with a conventional sinogram grid[3,18].

The TVS1-DROP scheme, which was similar to the TVS algorithms used in previous
210 studies[11,12,13], applied the projection operator cyclically until all blocks of the data set had been processed. Following this, the feasibility proximity was checked including all histories in the data set.

**Algorithm 2:** Cyclic total-variation superiorization with DROP (TVS1-DROP)

1. set $k = 0$
2. set $x^k = x_{FBP}$ the initial FBP reconstruction, and $\beta_k = 1$
3. repeat for 10 cycles
4.     set $s$ to a subgradient of $\phi$ at $x^k$
5.     if $\|s\| > 0$ set $v^k = -s / \|s\|$
6.     else set $v^k = s$
7.     set *continue* = *true*
8.     while *continue*
9.         set $y^k = x^k + \beta_k v^k$
10.         calculate the merit function (total variation) with Eq. (7), and if $\phi(y^k) \leq \phi(x^k)$
11.             apply sequentially $M$ times the projection operator $\mathbf{P}_{t(k)}$ to $y^k$ (Eq.(5))
12.             calculate the feasibility proximity with Eq. (6) using histories from all $M$ blocks, and if $Pr(\mathbf{P}_M y) < Pr(x^k)$
13.                 set $x^{k+1} = \mathbf{P}_M y$
14.                 set *continue* = *false*
15.             else set $\beta_k = \beta_k / 2$
16.         else set $\beta_k = \beta_k / 2$

17.  set $k = k + 1$

The TVS2-DROP scheme applied the projection operator to a single block only, before continuing to the feasibility proximity check, which was performed only with histories from the subsequent block. This was justified since each block was composed of an equal number of histories from each projection angle and was thus representative of the data set as a whole.

**Algorithm 3:** Block total-variation superiorization with DROP (TVS2-DROP)

1. set $k = 0$
2. set $x^k = x_{\text{FBP}}$ the initial FBP reconstruction, and $\beta_k = 1$
3. repeat for each block over 10 cycles
4.  set $s$ to a subgradient of $\phi$ at $x^k$
5.  if $\|s\| > 0$ set $v^k = -s / \|s\|$
6.  else set $v^k = s$
7.  set *continue* = *true*
8.  while *continue*
9.   set $y^k = x^k + \beta_k v^k$
10.   calculate the merit function (total variation) with Eq. (7), and if
$$\phi(y^k) \leq \phi(x^k)$$
11.   apply the projection operator $\mathbf{P}_{t(k)}$ to $y$ (Eq. (5))
12.   calculate the feasibility proximity with Eq. (6) using histories from the current block, and if $Pr(\mathbf{P}_{t(k)} y) < Pr(x^k)$
13.    set $x^{k+1} = \mathbf{P}_M y$
14.    set *continue* = *false*

255         15.                     else set $\beta_k = \beta_k / 2$

            16.                     else set $\beta_k = \beta_k / 2$

            17.         set $k = k + 1$

In an effort to reduce image reconstruction time, the TVS1- and TVS2-DROP schemes were further modified by eliminating the computationally expensive feasibility proximity check in step 260 12. The modified schemes without the feasibility proximity checking are referred to as TVS1-DROP* and TVS2-DROP* in the remainder of the paper.

**C. Proton CT Monte Carlo simulations**

The Geant4[19] Monte Carlo pCT simulation geometry described in previous work[5] was used as the basis for the current work. The detector system consisted of four proton tracking planes and a 265 crystal calorimeter (Fig. 1). The 30 × 30 × 0.04 cm 2D sensitive silicon tracking planes were assigned a spatial resolution of 100 μm. The calorimeter detector was a cesium iodide 32 × 32 × 10 cm rectangular prism with perfect energy resolution, i.e., sources of detector noise were neglected.

The virtual phantoms used to quantify spatial and density resolution are shown in Fig. 2(a) and 2(b), respectively. Both phantoms had a diameter of 16 cm and contained two materials, 270 equivalent in chemical composition and electron density to brain and cranial bone as defined by the International Commission on Radiological Protection (ICRP)[20]. The spatial resolution phantom contained an additional central rectangular prism structure, having a cross-section of (0.82 × 0.82) mm$^2$, equal to the reconstruction pixel size. The electron density of this structure was 20 times greater than the surrounding brain material but retained the same chemical composition.

275      The incident, monoenergetic protons had an energy of 200 MeV and formed a two-dimensional (2D) parallel-beam geometry. One hundred and eighty projections with 2-degree intervals were simulated for each phantom. For each projection angle, the position in each tracking plane and energy deposited in the calorimeter were recorded for 20,000 protons.

The Geant4 standard model of hadronic ionization was implemented. This model employed the Bethe-Bloch relationship for proton energies above 2 MeV, which covered our energy range of interest for protons traversing the phantom. For ionization energy loss, the standard Geant4 configuration involves calculation of the mean value in 100 steps evenly spaced logarithmically in kinetic energies from 1 keV to 100 TeV. However, studies have shown that this default configuration is not accurate enough for pCT applications[21]. Based on the suggestion of Heimann *et al.*[21] the binning was calculated from 1 keV to 500 MeV in 2000 steps. Low energy elastic and inelastic nuclear collision models were enabled.

Based on the simulated proton data (entry and exit coordinates and energy deposited in the calorimeter) a 2D image of each phantom was reconstructed with the algorithms described above. These calculations were carried out on a general purpose graphical processing unit (GPGPU) workstation. The workstation consisted of a quad-core central processing unit (CPU) and two NVIDIA® Tesla C1060 GPUs (NVIDIA Corporation, Santa Clara, CA, USA). The GPGPU code was written with the "C for CUDA" toolkit, the drivers for which are freely available from NVIDIA®. To enable parallel execution on a dual GPGPU system, multi-threaded coding was implemented on the host CPU. This was done using the OpenMP application programming interface[22]. Only inherently parallel parts of the iterative reconstruction were executed on the GPGPUs. Projections within a block (the sum in Eq. (5)) suit the parallelization criteria well, as individual projections within a block are independent of the result of the other projections. Thus, following the completion of a block projection on the GPGPU, the summed array was returned to the CPU for processing and the sequential portion of the block-iterative algorithm carried out.

**D. Performance analysis**

The performance of the different variants of reconstruction algorithms was compared by obtaining quantities for spatial resolution, density resolution, and relative RSP accuracy as described below.

Spatial resolution of the reconstructed images was quantified with the 2D modulation transfer function (MTF). The MTF is a measure of the signal transmission properties of the imaging system as a function of spatial frequency. For this measure, the point spread function (PSF) of the image of the central dense rectangular prism in Fig. 2(a) was used. Following reconstruction, a 2D FFT of a $16 \times 16$ pixel region of interest centered on the PSF was carried out. Making use of the axial symmetry of the phantom, the MTF was obtained in the region of interest by averaging the magnitude of the $x$ and $y$ axial components of the resulting spatial frequency representation of the image.

Low-contrast density resolution was assessed with the contrast discrimination function (CDF). This is an objective statistical analysis method for determining the minimum contrast required to discriminate an object of a given size from the surrounding tissue[23]. The CDF was calculated by dividing the reconstructed image of the uniform phantom into a grid of objects, ranging from $1 \times 1$ to $10 \times 10$ pixels in size. The standard deviation of the distribution of mean pixel values within the grid elements were used to determine the minimum contrast detectable with a given confidence level. For a 95% confidence level, the detectable density difference between the object of selected grid size and the background was defined as 3.29 standard deviations of the mean pixel value distribution.

The quantitative accuracy of reconstructed RSP values was determined using histogram analysis and defining the relative RSP error as

$$\varepsilon_n = \sum_j \left| x_j^{'} - x_j^n \right| / \sum_j \left| x_j^{'} \right|, \tag{9}$$

where $x'_j$ is the RSP in pixel *j* of the phantom and $x^n_j$ is the reconstructed RSP in pixel *j* after *n* cycles.

## III. RESULTS

### A. Qualitative comparison

Images of the two virtual phantoms reconstructed with standard DROP, TVS1-DROP* and TVS2-DROP* are shown in Fig. 3. The images reconstructed with these variants of the TVS scheme without the feasibility proximity checking had a smaller or equivalent minimum relative error when compared to images reconstructed with the feasibility proximity check (see next section and Fig. 4). It should be noted that the images shown in Fig. 3 correspond to the image obtained at the cycle of minimum relative error, which was cycle 3 for standard DROP, and cycle 10 for TVS1-DROP* and TVS2-DROP*. Qualitatively, it can be seen that the TVS2-DROP* scheme had the lowest noise level, probably due to the extra perturbation steps.

### B. Quantitative accuracy

Fig. 4 displays the relative error as a function of cycle number for all reconstruction schemes. The images reconstructed with the TVS1-DROP and TVS1-DROP* schemes, i.e., with and without the feasibility proximity check, were equivalent in terms of quantitative RSP accuracy, and the relative error followed a monotonically decreasing trend. The removal of the feasibility proximity check made no difference as the check condition was never violated in this case. On the other hand, the removal of the feasibility proximity check made a difference for the TVS2-DROP scheme. Fig. 4 demonstrates that including the feasibility proximity check led to a progressive increase of the relative error after reaching a minimum similar to the standard DROP algorithm. This can be explained by the fact that the reduced $\beta$ dampens the noise-reducing effect of the perturbation step. Thus, as the standard DROP algorithm diverges from a low relative error, so does the more stringent TVS approach. This occurs with TVS2-DROP but not TVS1-DROP because violations of the feasibility proximity condition were only observed with the former. Without the feasibility proximity check, the relative error of the TVS2-DROP* scheme followed a monotonically
350 decreasing trend within the 10 cycles. The minimum relative error within the first 10 cycles was 2.64% with standard DROP, 1.96% with TVS1-DROP and TVS1-DROP*, 1.64% with TVS2-DROP, and 1.55% with TVS2-DROP*. These differences are a direct result of the various degrees of noise in the images reconstructed with the different schemes.

The results presented were obtained with the data subdivided into 12 blocks. The
355 reconstruction algorithms were also run with 180 blocks, but the results were very similar and are, therefore, not shown.

Histograms of the images presented in the top row of Fig. 2 were created to analyze the mean reconstructed value of the brain and bone-equivalent regions. Gaussian distributions were fitted to the peaks to model reconstruction noise. All schemes reconstructed the same mean RSP value for
360 the brain and bone-equivalent regions, within peak-fitting uncertainty. Thus, the TVS perturbation schemes did not adversely affect the accuracy of the reconstructed values of these materials.

**C. Spatial resolution**

Due to their superior noise performance and reduced reconstruction time, further analysis was only performed for the TVS1-DROP* and TVS2-DROP* schemes, which were compared to
365 the standard DROP reconstruction scheme. The MTFs associated with each algorithm are plotted in Fig. 5. For any spatial frequency, the TVS1-DROP* scheme had larger MTF values and thus superior spatial resolution than the standard DROP scheme. The TVS1-DROP* and the TVS2-DROP* schemes performed similarly in terms of spatial resolution, with the TVS1-DROP scheme being marginally better. The improved spatial resolution with the TVS reconstruction schemes can
370 be attributed to the greater number of cycles being performed before reaching the lowest relative error. It has been observed previously that pCT spatial resolution improves with cycle number when employing an MLP formalism in conjunction with an iterative algorithm[3]. This is an important result since reconstruction algorithms that improve density resolution (see below) often display inferior spatial resolution.

### D. Density resolution

The CDFs associated with the standard DROP and the reduced TVS-DROP* schemes are plotted in Fig. 6. While the TVS1-DROP* scheme performed only slightly better than the standard DROP scheme, the TVS2-DROP* scheme performed much better than the other two algorithms. For objects as small as 1 mm$^2$, the TVS2-DROP* algorithm allowed contrast discrimination between 1% and 1.5%. The superior contrast discrimination of the TVS2-DROP* scheme can be attributed to the combination of reduced image noise and improved spatial resolution.

## IV. DISCUSSION

The new concept of superiorization, as outlined in Ref. 14, can be applied to inverse problems in which a reduction, which is not necessarily a minimization, of a given merit function subject to convex feasibility constraints is required. In this work, we have focused on the application of the general superiorization scheme to pCT reconstructions and made certain modifications to suit the task at hand. Central to the superiorization concept applied to pCT or other iterative image reconstruction methods is the act of perturbing the calculated image estimates between the iterative steps of a feasibility seeking projection method. By choosing the method of perturbation appropriately, significant beneficial alterations to the sequence of reconstructed images were achieved.

In this study two superiorization schemes, TVS1-DROP and TVS2-DROP, based on a reduction of the TV of the pCT image reconstructed with the DROP algorithm were investigated.

The two schemes differed in the frequency of perturbation per reconstruction cycle. TVS1-DROP performed only one perturbation in each cycle, while TVS2-DROP made use of a perturbation at each block iteration (12 per cycle in this case). Both TVS-DROP schemes were found to improve image quality relative to standard DROP. In particular, the additional perturbation steps utilized in TVS2-DROP resulted in the greatest reduction of image noise and superior density resolution.

Attention must be paid to the extra computation time when incorporating TVS schemes into pCT reconstruction algorithms. The calculation of the TV merit function (Eq. (7)) and the perturbation vector $v^k$ could increase image reconstruction time when the dimension of the image is large, but the main cause for the excess reconstruction time in both TVS-DROP schemes was the calculation of the feasibility proximity function (Eq. (6)). In the best case scenario, in which the feasibility proximity check is never violated, a minimum of two projection cycles must be carried out for each conventional DROP cycle. To counteract the increased computation time, the two TVS-DROP schemes were also executed without the feasibility proximity checking step, denoted by TVS1-DROP* and TVS2-DROP*. This innovation halved the reconstruction time of both TVS-DROP schemes and further reduced the image noise of TVS2-DROP, while having no detrimental effect on the other performance parameters when compared with TVS1-DROP.

The purpose of the feasibility proximity function, is to ensure that superiorization with respect to TV does not force the reconstructed image away from the measured data. However, due to inaccuracies in the forward and backprojection operator in the iterative DROP algorithm, which arise from multiple Coulomb scattering and energy straggling of protons when traversing the object, the residual is not an accurate guide to image quality. The results presented here suggest that the TVS2-DROP scheme suppresses the inconsistencies present in the measured data. This means that the high spatial resolution inherent in iterative algorithms employing the MLP formalism can be successfully combined with the low-contrast sensitivity due to TV superiorization.

Another key finding of our investigation is the improvement in spatial resolution measured with both TVS-DROP reconstruction schemes relative to the standard DROP approach. While the TVS1-DROP* scheme displayed a marginally superior spatial resolution than TVS2-DROP*, the latter still resulted in superior spatial resolution relative to the image reconstructed with the standard DROP reconstruction despite its better noise reduction. We have noticed that previous attempts to improve density resolution by "smoothing" the reconstructed image, in general, resulted in a degradation of spatial resolution. This is not the case with both TVS-DROP schemes, where the spatial resolution was maintained or improved.

In this work we have done a first investigation of the performance of TV-superiorization methods in terms of quantitative accuracy, spatial resolution and low contrast density resolution, with pCT data acquired from largely uniform virtual phantoms. TV-based methods are known to work well for such piecewise constant objects. Our work in progress with an experimental pCT system will provide additional opportunities to study the usefulness of TVS schemes in pCT image reconstruction of realistic anthropomorphic phantoms

Superiorization is a promising new paradigm that has already been successfully applied to X-ray CT (see Refs 11, 12, 13, and 14), particularly in conjunction with the TV cost function. Our current report is a "feasibility study" intended to show that the combination of TV and superiorization can be successfully translated from X-ray CT to pCT; two fundamentally different imaging techniques. This "opens the door" for testing of other functions in the superiorization methodology that have already been used as cost functions in the context of denoising via *optimization*, which may prove to be more powerful than TV.

## V. CONCLUSION

440  Superiorization is a general mathematical concept that was applied to pCT image reconstruction in this work. Two TVS schemes were applied as an add-on to the standard DROP reconstruction algorithm, which we had previously used in pCT image reconstruction. It was found that both spatial and density resolution were improved by both TVS-DROP schemes, while quantitative accuracy was maintained. To reduce reconstruction time, a costly step of feasibility
445  proximity checking was removed from the TVS-DROP schemes. This resulted in halving the computation time and in further improved image quality. Considering the significant low-contrast advantages of the TVS2-DROP* scheme, we plan to implement this scheme in our future pCT image reconstruction of pCT images obtained with an experimental pCT system.

## ACKNOWLEDGEMENTS

450  We thank the anonymous referees for their constructive comments. The first author was supported in part by Grant Number 08/RSA/1-02 from the Cancer Institute New South Wales and the Department of Radiation Medicine at Loma Linda University Medical Center, California. Yair Censor was supported by NIH grant R01HL070472 administered by the National Heart, Lung, And Blood Institute. The authors would like to thank Gabor Herman and Ran Davidi from the Graduate
455  Center of the City University of New York for their valuable advice on superiorization schemes.

# REFERENCES

[1] R.W. Schulte, V. Bashkirov, T. Li, Z. Liang, K. Mueller, J. Heimann, L.R. Johnson, B. Keeney, H.F.-W. Sadrozinski, A. Seiden, D.C. Williams, L. Zhang, Z. Li, S. Peggs, T. Satogata, and C. Woody, "Conceptual design of a proton computed tomography system for applications in proton radiation therapy", *IEEE Trans. Nucl. Sci.*, **51**, 866–872 (2004).

[2] T. Li, Z. Liang, J.V. Singanallur, T.J. Satogata, D.C. Williams and R.W. Schulte, "Reconstruction for proton computed tomography by tracing proton trajectories: A Monte Carlo study", *Med. Phys.*, **33**, 699–706 (2006).

[3] S.N. Penfold, R.W. Schulte, Y. Censor, V. Bashkirov, S. McAllister, K.E. Schubert, and A.B. Rosenfeld, "Block-iterative and string-averaging projection algorithms in proton computed tomography image reconstruction", in: *Biomedical Mathematics: Promising Directions in Imaging, Therapy Planning and Inverse Problems*, Y. Censor, M. Jiang, and G. Wang (Eds.), Medical Physics Publishing, Madison, WI, 347–368 (2010).

[4] R.W. Schulte, S.N. Penfold, J.T. Tafas and K.E. Schubert, "A maximum likelihood proton path formalism for application in proton computed tomography", *Med. Phys.*, **35**, 4849–4856 (2008).

[5] S.N. Penfold, R.W. Schulte, K.E. Schubert, A.B. Rosenfeld, "A more accurate reconstruction system matrix for quantitative proton computed tomography", *Med. Phys.*, **36**, 4511–4518 (2009).

[6] H.H. Baushke and J.M. Borwein, "On projection algorithms for solving convex feasibility problems", *SIAM Review*, **38**, 367–426 (1996).

[7] Y. Censor, S.A. Zenios, *Parallel Optimization: Theory, Algorithms, and Applications*, (Oxford University Press, New York, NY, 1997).

[8] P.L. Combettes and J. Luo, "An adaptive level set method for nondifferentiable constrained image recovery", *IEEE Trans. Image Proc.*, **11**, 1295–1304 (2002).

[9] P.L. Combettes and J.C. Pesquet, "Image restoration subject to a total variation constraint", *IEEE Trans. Image Proc.*, **13**, 1213–1222 (2004).

[10] E.Y. Sidky and X. Pan, "Image reconstruction in circular cone-beam computed tomography by constrained, total-variation minimization", *Phys. Med. Biol.*, **53**, 4777–4807 (2008).

[11] D. Butnariu, R. Davidi, G.T. Herman, and I.G. Kazantsev, "Stable convergence behavior under summable perturbations of a class of projection methods for convex feasibility and optimization problems", *IEEE Jour. Sel. Top. Sig. Proc.*, **1**, 540–547 (2007).

[12]G.T. Herman and R. Davidi, "Image reconstruction from a small number of projections", *Inv. Prob.*, **24**, online paper number 045011 (2008).

[13]R. Davidi, G.T. Herman, and Y. Censor, "Perturbation-resilient block-iterative projection methods with application to image reconstruction from projections", *Inter. Trans. Oper. Res.*, **16**, 505–524 (2009).

[14]Y. Censor, R. Davidi, and G.T. Herman, "Pertubation resilience and superiorization of iterative algorithms", *Inv. Prob.*, **26**, 065008 (2010).

[15]Y. Censor, T. Elfving, G.T. Herman, and T. Nikazad, "On diagonally-relaxed orthogonal projection methods", *SIAM Jour. Sci. Comp.*, **30**, 473–504 (2008).

[16]R. Aharoni and Y. Censor, "Block-iterative projection methods for parallel computation of solutions to convex feasibility problems", *Lin. Alg. Appl.*, **120**, 165–175 (1989).

[17]P.P.B. Eggermont, G.T. Herman, and A. Lent, "Iterative algorithms for large partitioned systems, with applications to image reconstruction", *Lin. Alg. Appl.*, **40**, 37–67 (1981).

[18]R.W. Schulte, V. Bashkirov, M.C. Klock, T. Li, A.J. Wroe, I. Evseev, D.C. Williams, T. Satogata, "Density resolution of proton computed tomography", *Med. Phys.*, **32**, 1035–1046 (2005).

[19]S. Agostinelli et al., "GEANT4—A simulation toolkit", *Nucl. Instr. Meth. Phys. Res. A*, **506**, 250–303 (2003).

[20]International Commission on Radiological Protection, "*Report of the Task Group on Reference Man*," ICRP Publication 23 (Elsevier, Philadelphia, 1975).

[21]J. Heimann, L. Johnson, T. Satogata, and D.C. Williams, "The requirements and limitations of computer simulations applied to proton computed tomography", *2004 IEEE Nucl. Sci. Symp. and Med. Imag. Conf. Rec.*, T. Seibert (Ed), 3663–3666 (2004).

[22] L. Dagum and R. Menon, "OpenMP: An Industry Standard API for Shared-Memory Programming", *IEEE Comp. Sci. Eng.*, **5**, 46–55 (1998).

[23]J. Hsieh, *Computed Tomography: Principles, Design, Artifacts and Recent Advances*, (SPIE, Bellingham, WA, 2003).

**FIGURES**

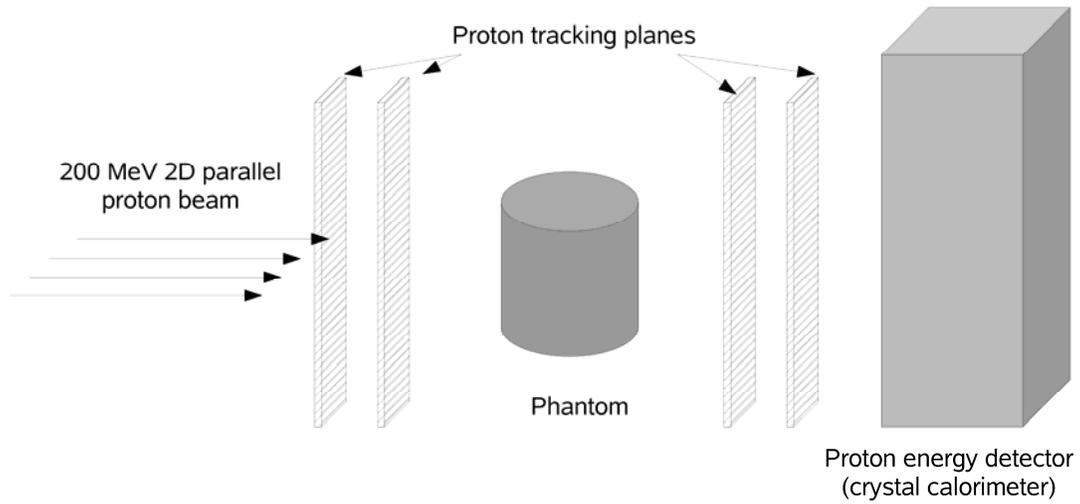

460 FIG. 1. Illustration of the Geant4 simulation geometry.

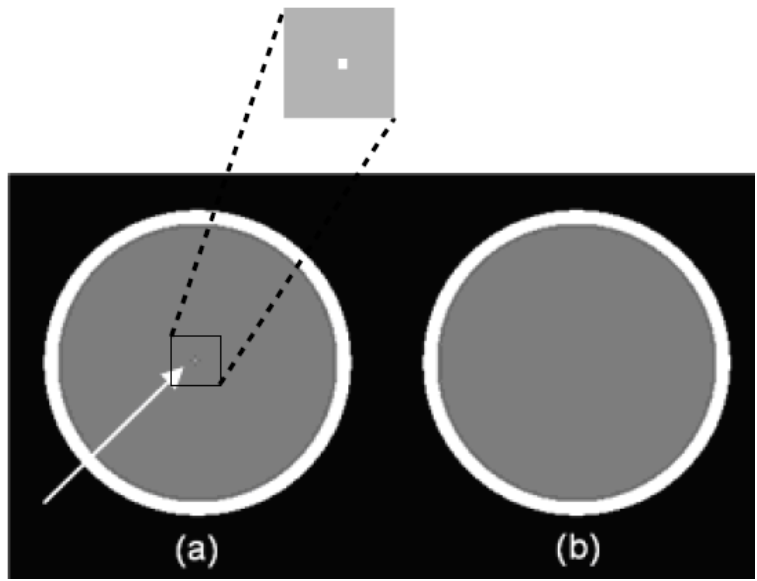

FIG. 2. Cross-sections of the cylindrical phantoms used in the Geant4 pCT simulations. (a) Phantom with central dense structure (indicated by arrow) to quantify spatial resolution. (b) Phantom with uniform interior. The white regions were assigned density and composition of bone and the gray regions density and composition of brain.

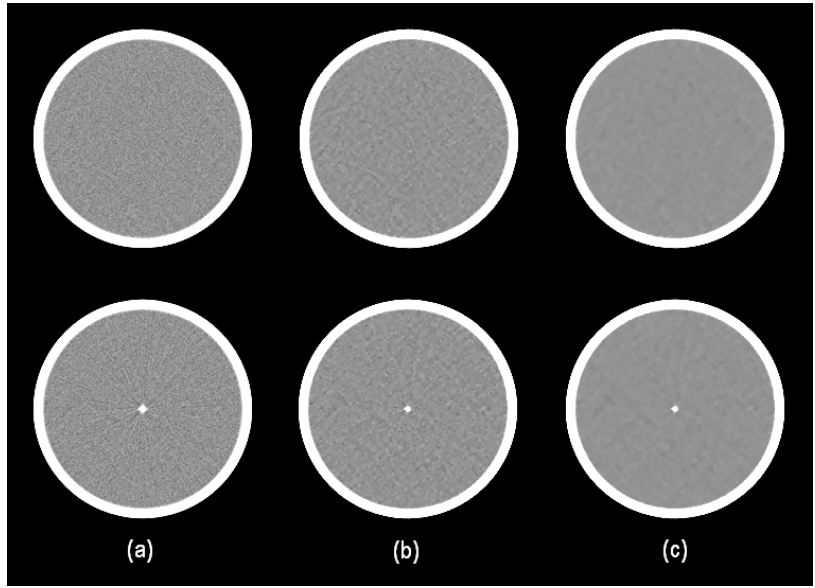

465

FIG. 3. Images reconstructed with (a) standard DROP, (b) TVS1-DROP*, and (c) TVS2-DROP*. Images in the top row are reconstructions of the uniform phantom and in the bottom are reconstructions of the spatial resolution phantom. The viewing window includes RSP values between 0.8 and 1.2.

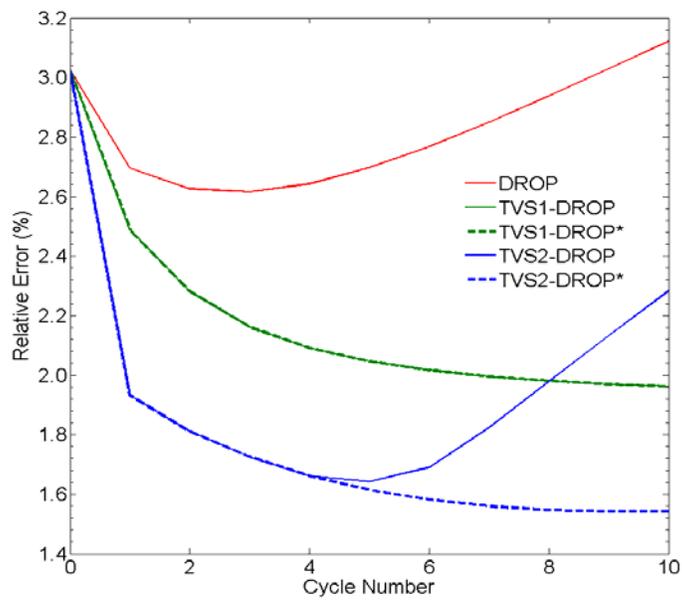

470 FIG. 4. Relative error as a function of cycle number for the various schemes. The relative error at cycle 0 corresponds to the relative error produced by the FBP algorithm, which was used to generate the initial point for the iterative TV-superiorization.

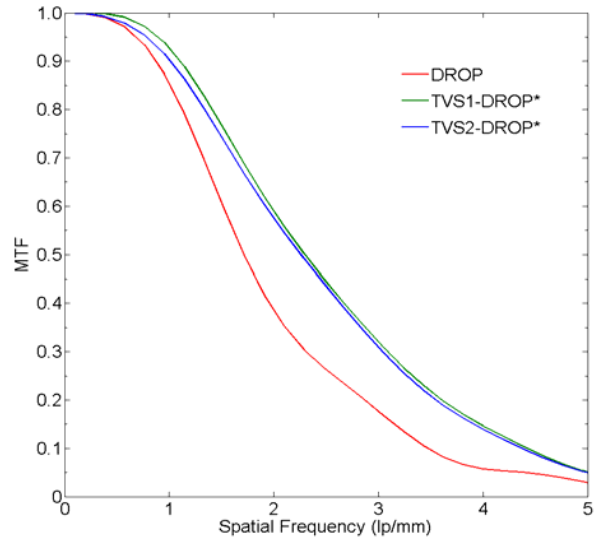

FIG. 5. MTF of the standard DROP and the reduced TVS-DROP* reconstruction schemes. The greater MTF value for
475  any give spatial frequency reflects the superior spatial resolution of the TVS schemes.

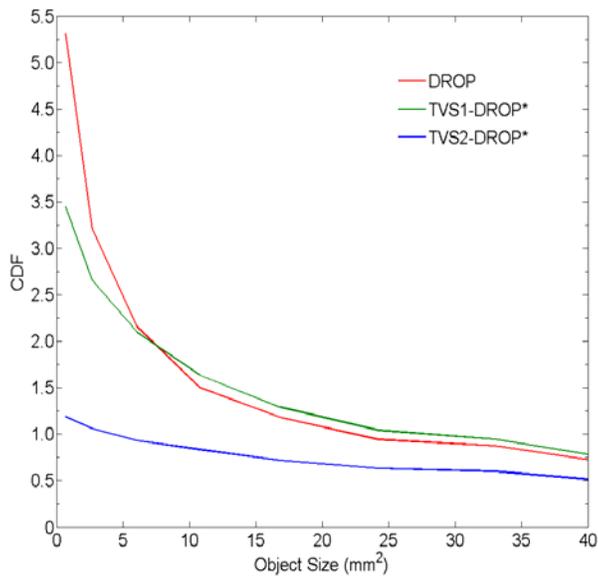

FIG. 6. CDF derived from the standard DROP and the reduced TVS-DROP* reconstruction schemes. The CDF specifies the percentage contrast required to discriminate an object of a give size from background with a 95% confidence level.